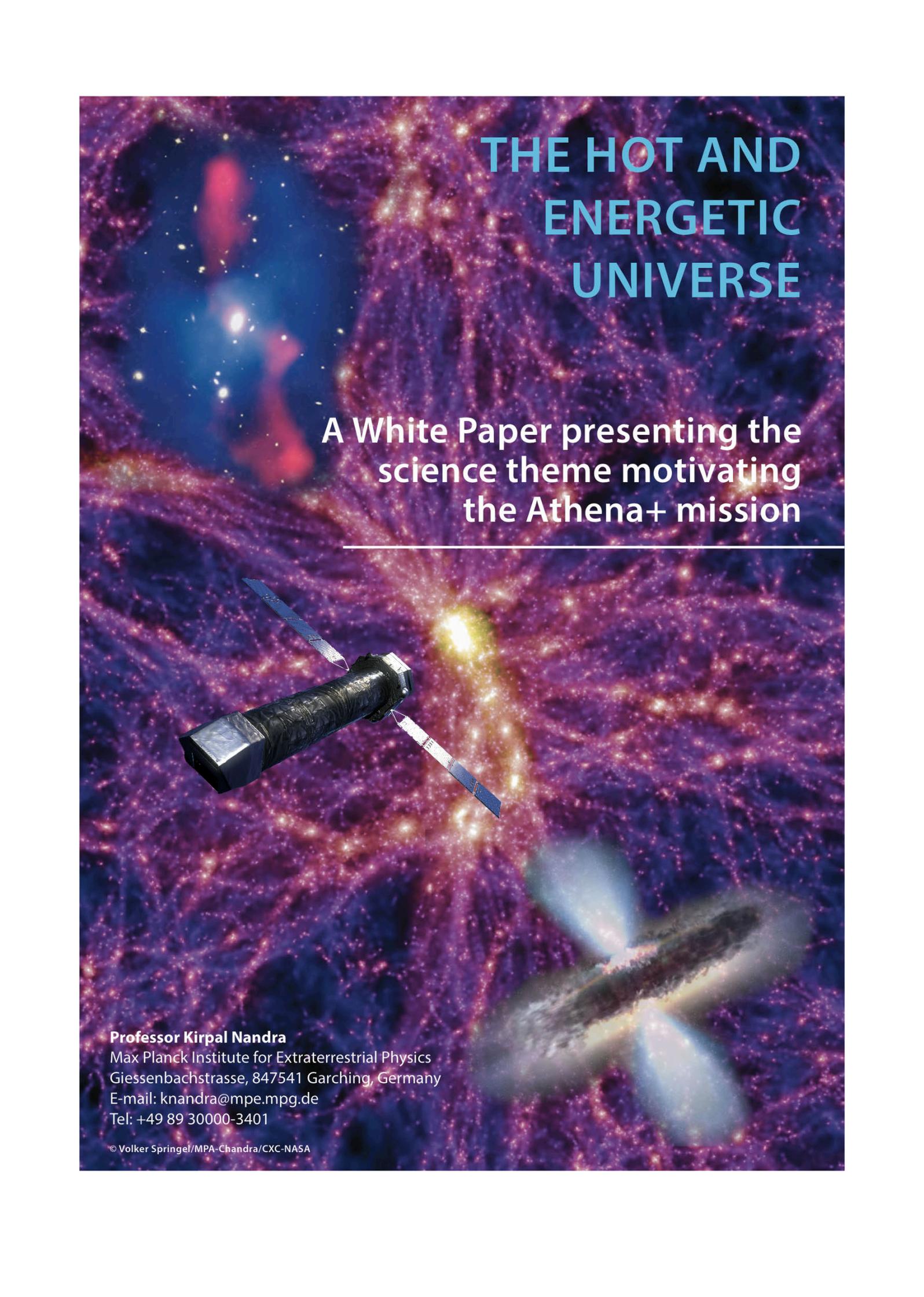

# THE HOT AND ENERGETIC UNIVERSE

## A White Paper presenting the science theme motivating the Athena+ mission

**Professor Kirpal Nandra**
Max Planck Institute for Extraterrestrial Physics
Giessenbachstrasse, 847541 Garching, Germany
E-mail: knandra@mpe.mpg.de
Tel: +49 89 30000-3401

# AUTHORS AND CONTRIBUTORS

**The Athena+ Co-ordination Group:** Xavier Barcons (ES), Didier Barret (FR), Andy Fabian (UK), Jan-Willem den Herder (NL), Kirpal Nandra (DE), Luigi Piro (IT), Mike Watson (UK)

**The Athena+ Working Groups[1]:** Christophe Adami (FR), **James Aird (UK),** Jose Manuel Afonso (PT), Dave Alexander (UK), Costanza Argiroffi (IT), Lorenzo Amati (IT), Monique Arnaud (FR), Jean-Luc Atteia (FR), Marc Audard (CH), Carles Badenes (US), Jean Ballet (FR), Lucia Ballo (IT), Aya Bamba (JP), Anil Bhardwaj (IN), Elia Stefano Battistelli (IT), Werner Becker (DE), Michaël De Becker (BE), Ehud Behar (IL), Stefano Bianchi (IT), Veronica Biffi (IT), Laura Bîrzan (NL), Fabrizio Bocchino (IT), Slavko Bogdanov (US), Laurence Boirin (FR), Thomas Boller (DE), Stefano Borgani (IT), Katharina Borm (DE), Nicolas Bouché (FR), Hervé Bourdin (IT), Richard Bower (UK), Valentina Braito (IT), Enzo Branchini (IT), **Graziella Branduardi-Raymont (UK),** Joel Bregman (US), Laura Brenneman (US), Murray Brightman (DE), Marcus Brüggen (DE), Johannes Buchner (DE), Esra Bulbul (US), Marcella Brusa (IT), Michal Bursa (CZ), Alessandro Caccianiga (IT), Ed Cackett (US), Sergio Campana (IT), Nico Cappelluti (IT), **Massimo Cappi (IT), Francisco Carrera (ES),** Maite Ceballos (ES), Finn Christensen (DK), You-Hua Chu (US), Eugene Churazov (DE), Nicolas Clerc (DE), Stephane Corbel (FR), Amalia Corral (ES), **Andrea Comastri (IT), Elisa Costantini (NL), Judith Croston (UK),** Mauro Dadina (IT), Antonino D'Ai (IT), **Anne Decourchelle (FR),** Roberto Della Ceca (IT), Konrad Dennerl (DE), Klaus Dolag (DE), **Chris Done (UK), Michal Dovciak (CZ),** Jeremy Drake (US), Dominique Eckert (CH), Alastair Edge (UK), **Stefano Ettori (IT),** Yuichiro Ezoe (JP), Eric Feigelson (US), Rob Fender (UK), Chiara Feruglio (FR), **Alexis Finoguenov (FI),** Fabrizio Fiore (IT), Massimiliano Galeazzi (IT), Sarah Gallagher (CA), Poshak Gandhi (UK), Massimo Gaspari (IT), Fabio Gastaldello (IT), **Antonis Georgakakis (DE),** Ioannis Georgantopoulos (GR), Marat Gilfanov (DE), Myriam Gitti (IT), Randy Gladstone (US), Rene Goosmann (FR), Eric Gosset (BE), Nicolas Grosso (FR), Manuel Guedel (AT), Martin Guerrero (ES), Frank Haberl (DE), Martin Hardcastle (UK), Sebastian Heinz (US), Almudena Alonso Herrero (ES), Anthony Hervé (FR), Mats Holmstrom (SE), Kazushi Iwasawa (ES), **Peter Jonker (NL), Jelle Kaastra (NL),** Erin Kara (UK), Vladimir Karas (CZ), Joel Kastner (US), Andrew King (UK), Daria Kosenko (FR), Dimita Koutroumpa (FR), Ralph Kraft (US), Ingo Kreykenbohm (D), Rosine Lallement (FR), Giorgio Lanzuisi (GR), J. Lee (US), Marianne Lemoine-Goumard (FR), Andrew Lobban (UK), Giuseppe Lodato (IT), Lorenzo Lovisari (DE), Simone Lotti (IT), Ian McCharthy (UK), Brian McNamara (CA), Antonio Maggio (IT), Roberto Maiolino (UK), Barbara De Marco (DE), Domitilla de Martino (IT), Silvia Mateos (ES), **Giorgio Matt (IT),** Ben Maughan (UK), Pasquale Mazzotta (IT), Mariano Mendez (NL), Andrea Merloni (DE), Giuseppina Micela (IT), Marco Miceli (IT), Robert Mignani (IT), Jon Miller (US), Giovanni Miniutti (ES), Silvano Molendi (IT), Rodolfo Montez (ES), Alberto Moretti (IT), **Christian Motch (FR),** Yaël Nazé (BE), Jukka Nevalainen (FI), Fabrizio Nicastro (IT), Paul Nulsen (US), Takaya Ohashi (JP), **Paul O'Brien (UK),** Julian Osborne (UK), Lida Oskinova (DE), Florian Pacaud (DE), Frederik Paerels (US), Mat Page (UK), Iossif Papadakis (GR), Giovanni Pareschi (IT), Robert Petre (US), Pierre-Olivier Petrucci (FR), Enrico Piconcelli (IT), Ignazio Pillitteri (IT), C. Pinto (UK), Jelle de Plaa (NL), **Etienne Pointecouteau (FR),** Trevor Ponman (UK), Gabriele Ponti (DE), Delphine Porquet (FR), Ken Pounds (UK), **Gabriel Pratt (FR),** Peter Predehl (DE), Daniel Proga (US), Dimitrios Psaltis (US), David Rafferty (NL), Miriam Ramos-Ceja (DE), Piero Ranalli (IT), Elena Rasia (US), Arne Rau (DE), **Gregor Rauw (BE),** Nanda Rea (IT), Andy Read (UK), James Reeves (UK), **Thomas Reiprich (DE),** Matthieu Renaud (FR), Chris Reynolds (US), Guido Risaliti (IT), Jerome Rodriguez (FR), Paola Rodriguez Hidalgo (CA), Mauro Roncarelli (IT), David Rosario (DE), Mariachiara Rossetti (IT), Agata Roszanska (PL), Emmanouil Rovilos (UK), Ruben Salvaterra (IT), Mara Salvato (DE), Tiziana Di Salvo (IT), **Jeremy Sanders (DE),** Jorge Sanz-Forcada (ES), Kevin Schawinski (CH), Joop Schaye (NL), Axel Schwope (D), **Salvatore Sciortino (IT),** Paola Severgnini (IT), Francesco Shankar (FR), Debora Sijacki (UK), Stuart Sim (IE), Christian Schmid (DE), Randall Smith (US), Andrew Steiner (US), Beate Stelzer (IT), Gordon Stewart (UK), Tod Strohmayer (US), Lothar Strüder (DE), Ming Sun (US), Yoh Takei (JP), V. Tatischeff (FR), Andreas Tiengo (IT), Francesco Tombesi (US), Ginevra Trinchieri (IT), T.G. Tsuru (JP), Asif Ud-Doula (US), Eugenio Ursino (NL), Lynne Valencic (US), Eros Vanzella (IT), Simon Vaughan (UK), Cristian Vignali (IT), Jacco Vink (NL), Fabio Vito (IT), Marta Volonteri (FR), Daniel Wang (US), Natalie Webb (FR), Richard Willingale (UK), **Joern Wilms (DE),** Michael Wise (NL), Diana Worrall (UK), Andrew Young (UK), Luca Zampieri (IT), Jean In't Zand (NL), Silvia Zane (UK), Andreas Zezas (GR), Yuying Zhang (DE), Irina Zhuravleva (US).

*The Japanese X-ray astronomy community, as represented by the High-energy Astrophysics Association in Japan (HEAPA) is pleased to indicate its endorsement and support for the selection of the "Hot and Energetic Universe" as the science theme for an ESA large mission in 2028, to be implemented via an X-ray observatory.*

---

[1] Bold face for Working Group chairs. List of the 1100+ *Athena*+ supporters see: http://www.the-athena-x-ray-observatory.eu

# 1. EXECUTIVE SUMMARY

This White Paper advocates the need for a transformational leap in our understanding of two key questions in astrophysics: 1) How does ordinary matter assemble into the large scale structures that we see today? and 2) How do black holes grow and shape the Universe?

To understand the first of these questions, we must determine the physical evolution of clusters and groups of galaxies from their formation epoch at $z \sim 2$-3 to the present day. These structures grow over cosmic time by accretion of gas from the intergalactic medium, with the endpoint of their evolution being today's massive clusters of galaxies, the largest bound structures in the Universe. Hot gas in clusters, groups and the intergalactic medium dominates the baryonic content of the local Universe, so understanding how this component forms and evolves is a crucial goal.

While the framework for the growth of structure is set by the large scale dark matter distribution, processes of an astrophysical origin also have a major effect. To understand them, it is necessary to measure the velocities, thermodynamics and chemical composition of the gas to quantify the importance of non-gravitational heating and turbulence in the structure assembly process. The temperature of the hot gas is such that it emits copiously in the X-ray band, but current and planned facilities do not provide sufficient collecting area and spectral resolution to settle the issue of how ordinary matter forms the large scale structures that we see today. The key breakthrough is to enable spectroscopic observations of clusters beyond the local Universe, out to $z = 1$ and beyond, and spatially resolved spectroscopy to map the physical parameters of bound baryonic structures. Technological advances in X-ray optics and instrumentation can deliver simultaneously a factor 10 increase in both telescope throughput and spatial resolving power for high resolution spectroscopy, allowing the necessary physical diagnostics to be determined at cosmologically relevant distances for the first time.

One of the critical processes shaping hot baryon evolution is energy input – commonly known as feedback – from supermassive black holes. Remarkably, processes originating at the scale of the black hole event horizon seem able to influence structures on scales 10 orders of magnitude larger. This feedback is an essential ingredient of galaxy evolution models, but it is not well understood. X-ray observations are again the key to further progress, revealing the mechanisms which launch winds close to black holes and determining the coupling of the energy and matter flows on larger galactic and galaxy cluster scales.

The widespread importance of black hole feedback means that we cannot have a complete understanding of galaxies without tracking the growth of their central supermassive black holes through cosmic time. A key goal is to push the frontiers of black hole evolution to the redshifts where the first galaxies are forming, at $z = 6$-10. X-ray emission is the most reliable and complete way of revealing accreting black holes in galaxies, but survey capabilities need to be improved by a factor $\sim 100$ over current facilities to reach these early epochs and perform a census of black hole growth. This requires a combination of high sensitivity, which in turn depends on large throughput and good angular resolution, and wide field of view. Again, the required technologies to provide this leap in wide field X-ray spectral imaging are now within our grasp. The same high throughput needed to detect these early black holes will also yield the first detailed X-ray spectra of accreting black holes at the peak of galaxy growth at $z = 1$-4, measurements which are impossible with current instrumentation. These spectra will show, for example, if the heavily obscured phase of black hole evolution is associated with the termination of star formation in galaxies via feedback.

These topics comprise the major elements of the science theme **The Hot and Energetic Universe** (Figure 1). The Advanced Telescope for High-energy Astrophysics (hereafter ***Athena***+)[2] mission provides the necessary angular resolution, spectral resolution, throughput, detection sensitivity, and survey grasp to revolutionize our understanding of these issues. These capabilities will also provide a powerful observatory to be used in all areas of astrophysics. The technologies for the mission are mature, being based on much previous heritage and major technology developments. A lightweight X-ray telescope based on ESA's Silicon Pore Optics (SPO) technology provides large effective area with excellent angular resolution, combining with state-of-the-art instrumentation for spatially resolved high resolution X-ray spectroscopy (provided by the X-ray Integral Field Unit, X-IFU) and wide field X-ray imaging (provided by the Wide Field Imager, WFI). *Athena*+ will open up a vast discovery space leading to completely new areas of scientific investigation, continuing the legacy of discovery that has characterized X-ray astronomy since its inception. The implementation of *Athena*+ for launch in 2028 will guarantee a transformation in our understanding of **The Hot and Energetic Universe**, and establish European leadership in high-energy astrophysics for the foreseeable future.

---

[2] *Athena*+ as the successor of Athena, considered for L1, but with enhanced capabilities in terms of angular resolution, effective area, and instrument fields of view.

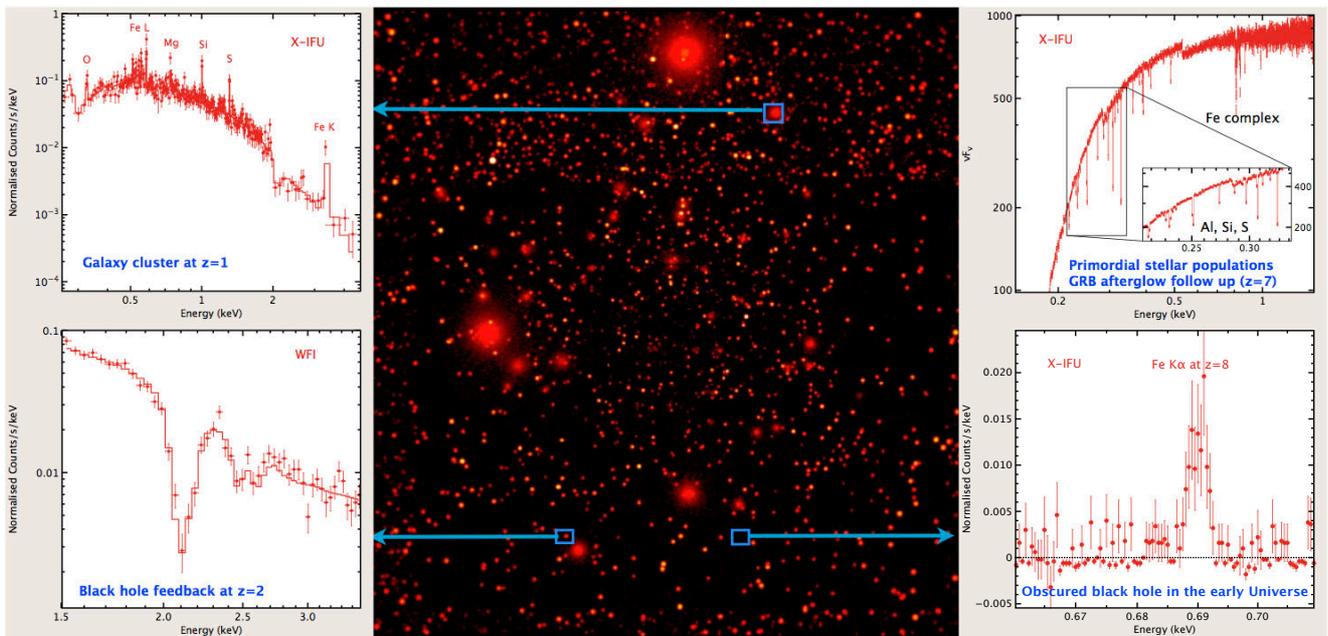

Figure 1: *Athena+* will provide revolutionary advances in our knowledge of the Hot and Energetic Universe. The central panel is a simulated deep WFI observation, while the four surrounding spectra illustrate advances in different science areas, none of which are possible with current facilities.

## 2. SCIENCE THEME: THE HOT AND ENERGETIC UNIVERSE

### 2.1. The Hot Universe: How does ordinary matter assemble into the large scale structures that we see today?

In the past decade or more, major observational and theoretical resources have been focused on the understanding of the formation and evolution of galaxies, and this will continue in the future as major facilities are implemented. A significant fraction of galaxies are trapped in larger scale structures – groups, clusters– whose baryonic content is nonetheless completely dominated by hot gas, with stars accounting for less than 15%. In fact, hot gas may dominate the total baryonic content of the local Universe. Groups and clusters of galaxies, as the next step up in the hierarchy of the Universe from galaxies themselves, are fundamental components of the Universe. While the backbone of the large scale structure of the Universe is determined by its cosmological parameters and by the gravitational interaction of the dominant dark matter, the assembly and evolution of baryonic structures are strongly affected by processes of astrophysical origin, which are often poorly known. A complete understanding of **The Hot Universe** - the baryonic gas that traces the most massive structures and drives the formation of galaxies within them - is a fundamental requirement of theories of structure formation and can only be achieved via X-ray observations. **Major astrophysical questions include:**

- How do baryons in groups and clusters accrete and dynamically evolve in the dark matter haloes?
- What drives the chemical and thermodynamic evolution of the Universe's largest structures?
- What is the interplay of galaxy, supermassive black hole, and intergalactic gas evolution in groups and clusters?
- Where are the missing baryons at low redshift and what is their physical state?

#### 2.1.1. The formation and evolution of groups and clusters of galaxies

In the $\Lambda$CDM cosmology, the first dark matter haloes are seeded by density fluctuations in the early Universe. These haloes accrete primordial gas and grow over cosmic time via hierarchical gravitational collapse. This process heats the gas to X-ray temperatures. Lying at the nodes of the cosmic web in today's highly structured Universe, galaxy clusters are the final product of this process. Over 80% of their total mass is in the form of dark matter. The remainder is composed of baryons trapped in the dark matter potential well, around 85% of which is diffuse, hot, metal-enriched, X-ray emitting plasma of the intra-cluster medium (ICM). The radiation from this gas and from the member galaxies reveals the interplay between the dark matter, the hot ICM and the cold baryons locked in stars and interstellar medium (e.g., Kravtsov & Borgani 2012).

By 2028, *Euclid* and LSST, and advances in numerical modelling, should have shown how dark matter structures assembled. However, X-ray observations are needed to understand the evolution of the baryons in the dark matter potential. Key observables include the gas temperature, abundance, velocity and ionisation stage, all of which are provided uniquely via observations of the X-ray continuum and emission lines. These observables reveal how gravitational energy shapes cluster assembly, showing how it is converted into thermal and non-thermal components in the ICM, and generates turbulence and kpc-scale bulk motions (e.g., Vazza et al. 2011). These processes have yet to be observed conclusively at the relevant spatial scales. Moreover, the contribution of the non-thermal component of the energy budget over time, and its influence on the formation and intrinsic properties of galaxy groups and clusters, cannot be determined with current and planned instruments. The spectral capabilities of the *Athena+* X-IFU will allow measurement of the projected ICM velocity field to the required precision through determination of spectral line broadening due to turbulence (see Figure 2) and measurements of line shift due to bulk motion. Fully constraining these basic astrophysical processes will determine the large scale properties of the ICM for nearby clusters, the first step in revealing how the baryons evolve in dark matter structures.

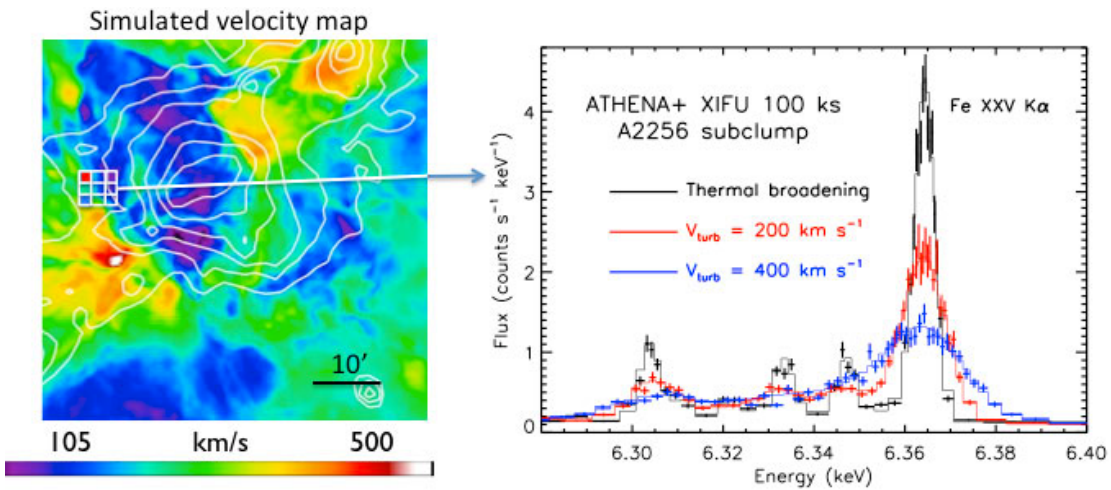

Figure 2: *Athena+* **X-IFU spectrum of a subclump in the galaxy cluster A2256, demonstrating the high precision measurements possible for the ICM velocity field.** *Left*: Velocity map of a cosmological hydrodynamical simulation of a perturbed galaxy cluster of about $M_{200} \sim 10^{15}$ $M_\odot$ with X-ray surface brightness contours overlaid. *Right:* Simulated spectrum for a 100 ks observation with *Athena+* X-IFU for a 1.5 arcmin. region (one of the 9 small regions shown on the image), showing the turbulent broadening of the Fe XXV Kα line . Simulated data with $v_{turb}$ =200 km s$^{-1}$ are shown in red. Black and blue represent the model with $v_{turb}$=0 and $v_{turb}$ =400 km s$^{-1}$, respectively. For an input turbulent velocity of 0, 200, 400 km s$^{-1}$, the 1σ statistical uncertainty is ±20, ±5, ±10 km s$^{-1}$, respectively.

The interplay of the cluster member galaxies with the ICM can further modify its properties beyond the simple expectations from pure gravitational collapse. For instance, supernova winds can add energy and eject enriched gas into the cluster atmosphere. It is also becoming clear that supermassive black holes inject enough energy to affect the ICM on Mpc scales. Cooling of the denser parts of the ICM feeds the central black hole, resulting in a feedback cycle. The history of this feedback over cosmic time is unknown, but can be quantified via the measurement of the entropy distribution and its evolution (e.g., Ettori et al. 2004). Entropy can be directly derived only from measurements of X-ray surface brightness and gas temperature. The *Athena+* X-IFU and WFI will measure the gas entropy profiles for groups and clusters on all mass scales out to $z \sim 2$. This will determine the non-gravitational energy input over the whole volume from the centre to the outskirts back to the epoch when star formation and accretion activity - and hence feedback processes - were most active.

Clusters are still forming today, and the cluster outskirts, extending across the virial radius and occupying about 85% of the cluster volume, are thus expected to be undergoing strong energetic activity as material is accreted into the dark matter potential (e.g., Reiprich et al. 2013). These are the regions where energy is first transferred into the ICM through merging events. This process will be resolved by *Athena+*. WFI and X-IFU observations will allow routine mapping of the outskirts of nearby clusters, providing their emission measure, temperature and metallicity. This will reveal the physical state of this accreting material, leading to a complete picture of the structure assembly process.

*Athena+* will probe the evolution of the scaling relations (e.g., Giodini et al., 2013) that link the total mass of a cluster and its observable properties across a wide range of masses up to $z\sim2$. Competing feedback and energy input models, which are degenerate at low redshift, will be constrained for the first time at high redshift (e.g., Short et al. 2010). Finally, while *eROSITA* will detect the X-ray emission of all massive clusters to $z\sim1.5$, much deeper X-ray observations are required to reveal low mass galaxy groups out to the formation epoch. Detection of an extended hot ICM is the unmistakable proof that a galaxy group is fully collapsed and possesses a deep gravitational potential well. The sensitivity and large field of view of the *Athena+* WFI will open up a discovery space for baryonic structures in their infancy at $z>2$ (e.g., Gobat et al., 2011). *Athena+* observations provide not only the means to discover these objects, but also measurements of the physical properties needed to understand how they formed.

### 2.1.2. The chemical history of hot baryons

Elements like carbon and beyond are generated via stellar processes, and form the basis of everything around us. Much effort has been devoted to understanding how these "metals" are generated and their distribution within galaxies. On the other hand, metals can easily be expelled from galaxies, for example by the action of supernovae and AGN winds. Clusters are ideal laboratories to study the production and distribution of heavy metals, having been enriched by member galaxies throughout their lifetime (e.g., de Plaa et al. 2007, Werner et al. 2008). In addition the ICM contains as many baryons as all the stars in the Universe. The metals enter the ICM via the gravitational action of ram-pressure stripping of infalling galaxies, merger-induced gas sloshing and galaxy-galaxy interactions, and feedback from super-winds in starburst galaxies and from AGN (e.g., Schindler & Diaferio, 2008, Böhringer & Werner 2010). How the gas and metals mix depends on transport processes in the magnetized ICM, which are currently poorly understood. X-ray observations of the lines emitted by the hot ICM are the only way to access information on their abundance and to probe its evolution to high redshift.

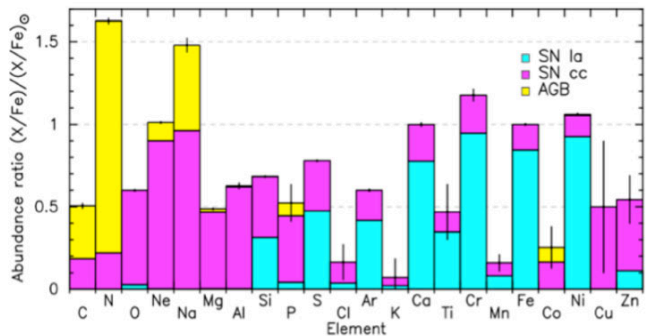

**Figure 3: Abundance measurements for a typical cluster of galaxies (AS 1101, 100 ks), illustrating the power of high precision *Athena+* X-IFU observations.** The expected abundance ratios relative to solar are shown for SNIa, SNcc, and AGB stars. Abundance measurements will strongly constrain the origin of the metals, the IMF and thus the star formation history.

With *Athena+* we will observe the epoch when the metals produced in the galaxies are ejected and redistributed in the ICM. The X-IFU will map the metal distribution and explore its relation to the metal ejection and transport processes, catching enrichment processes in the act. Radial profiles of the most abundant elements (O, Ne, Mg, Si, S) out to unprecedented distances will further show how elements are being distributed in the entire cluster volume. Looking at the metal distribution in the cluster outskirts, *Athena+* will determine the role of AGN feedback in shaping the metallicity profiles through expulsion of pre-enriched gas over cosmic time. *Athena+* will also determine the historical origins of ICM enrichment. Each source of metals (e.g., SNIa, SNcc, AGB stars) synthesizes heavy elements in different proportions, so their relative role can be assessed using abundance ratio measurements (e.g., of O/Fe and Si/Fe, Figure 3). *Athena+* will determine the main source of C and N, which can originate from a wide variety of sources including stellar mass loss from intermediate mass stars, and whose cosmic history is poorly known. Abundances of trace elements such as Cr and Mn, widely accessible for the first time, depend on the initial metallicity of the SNIa progenitor system, while N and Na determine the AGB star contribution.

Finally, current instruments have provided hints about cluster Fe abundance evolution from $z=1$ to the present (e.g., Baldi et al. 2012). Observations suggest that about half of the metals found in the ICM were released into the IGM and ICM prior to $z\sim1$. *Athena+* will measure the evolution of the most abundant elements with redshift, and their ratios, with unprecedented precision, tracing chemical evolution over cosmic time, telling the full story of how, where, and when the ICM was enriched. This is directly related to the metals that have been lost from the cluster member galaxies, so these observations reveal an important aspect of the evolution of galaxies in high-density environments.

### 2.1.3. Feedback in clusters of galaxies

The mechanical energy carried by jets from central AGN is now believed to control hot-gas cooling in massive ellipticals and groups and clusters of galaxies via a feedback loop in which jets heat the hot gas, suppressing star formation and regulating their own fuel supply. Current X-ray observations have revealed compelling evidence for such AGN feedback (e.g. McNamara & Nulsen 2007). In detail, however, there is a complex interplay between cooling and heating. The physics of how the balance between these processes is established and maintained, and how it evolves with time, is poorly understood. *Athena+* will make the first kinematic measurements on small spatial scales of the hot gas in galaxy, group and cluster haloes as it absorbs the impact of AGN jets. Combined with vastly improved ability to map thermodynamic conditions on angular scales well-matched to the jets, lobes and gas disturbances produced by them, this will provide answers to two key outstanding questions: how energy input from jets is dissipated and distributed throughout the ICM, and how the energy balance between cooling and heating is maintained and established in regions where the most massive galaxies are being formed.

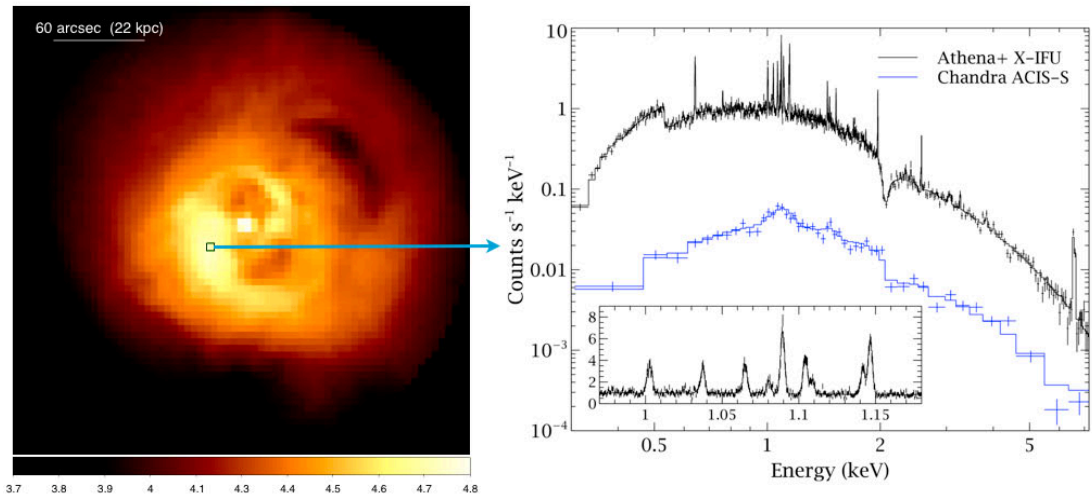

Figure 4: **Simulated *Athena+* observations of the Perseus cluster, highlighting the advanced capabilities for revealing the intricacies of the physical mechanisms at play.** The left panel shows a simulated 50ks X-IFU observation (0.5-7 keV), displayed on a log scale. The spectrum on the right is from the single 5"×5" region marked by the box, with the existing *Chandra* ACIS spectrum for comparison. The inset shows the region around the iron L complex. With such observations velocity broadening is measured to 10-20 km s$^{-1}$, the temperature to 1.5% and the metallicity to 3% on scales <10kpc in 20-30 nearby systems, and on <50kpc scales in hundreds of clusters and groups. Such measurements will allow us to pinpoint the locations of jet energy dissipation, determine the total energy stored in bulk motions and weak shocks, and test models of AGN fuelling so as to determine how feedback regulates hot gas cooling.

The X-IFU will map velocity structures and gas conditions on kpc-scales in the cores of galaxy groups and clusters where feedback is regulating cooling. The locations of heating and cooling will be pinpointed for the first time and the energy dissipation determined (Figure 4). With the WFI it will be possible to carry out the first population studies of the AGN-induced ripples, disturbances and weak shocks that are assumed to distribute the jet energy isotropically (e.g. Fabian et al. 2003), relating the mechanical energy stored in these disturbances, and its subsequent dissipation, to the environmental and AGN properties across a wide mass range. In addition to establishing the microphysics of AGN heating for the first time, *Athena+* also has the potential to determine how the AGN fuelling process is linked to the thermodynamical properties of the hot gas that absorbs the jet energy input, as is required if a self-regulated feedback process operates to suppress gas cooling and star formation. With an efficiency more than two orders of magnitude higher than the *XMM-Newton* RGS, the X-IFU will determine the gas cooling rates across a wide temperature range on spatial scales matched to the filamentary nebulae of cooler material observed to coincide with the regions of strongest X-ray cooling (e.g. Crawford et al. 1999). X-IFU measurements of the dynamics of the hot gas in the vicinity of the filaments will establish whether their motions are correlated, and distinguish locations where filaments are evaporated by the hot gas, where gas is thermally unstable to cooling and where mixing and possibly charge exchange are occurring, thus determining the relative importance of these processes. The role of AGN-induced turbulence in seeding thermal instabilities will be investigated via population studies, and robust jet power estimates from total mechanical energy input can be compared with accretion rates from hot and cold accretion models for the first time.

Beyond the core region, the energetic impact of radio jets and their role in building up entropy in group and cluster gas is poorly understood. The energy input from strong shocks expected to occur in typical environments is not taken into account in the scaling relations between radio luminosity and jet power (e.g. Bîrzan et al. 2008), and cannot be reliably determined from radio data. *Athena+* will enable the dynamics and source age and thus jet power to be assessed robustly via direct bulk velocity measurements of expanding hot gas shells around radio lobes extending up to Mpc scales. At higher redshifts, the identification of characteristic features associated with strong shocks in high-resolution WFI temperature maps will measure age, power and energetic impact for large representative samples.

### 2.1.4.    The missing baryons and the Warm-Hot Intergalactic Medium

The intergalactic medium contains 90% of the baryons at the current epoch, and is the visible tracer of the large scale dark matter structure of the local Universe. Theory predicts that the state of most of these baryons evolves from low temperatures, as manifested in the Lyα forest at $z>2$, to a warm-hot phase ($10^5$-$10^7$ K) at later times shaped by the filamentary structure of dark matter (Cen & Ostriker 2006). Most of the metals are predicted to reside in the warm-hot phase already at $z\sim4$. Thermal continuum emission from this gas is extremely hard to detect. The only characteristic radiation from this medium will be in the discrete transitions of highly ionized metals. Evidence for the warm tail of the WHIM, where 10-15% of the missing baryons reside, has been obtained via UV-absorption line studies with FUSE and HST-COS (Shull et al 2012). However, around 50% of the baryons at redshift $z<2$ and 90% of the metals at redshifts $z<3$, locked in the hot phase, remain unobserved. In order to reveal the underlying mechanisms driving the distribution of this gas on various scales, as well as different metal circulation and feedback processes the chemical and physical states of about a hundred filaments must be characterized. This can only be done in

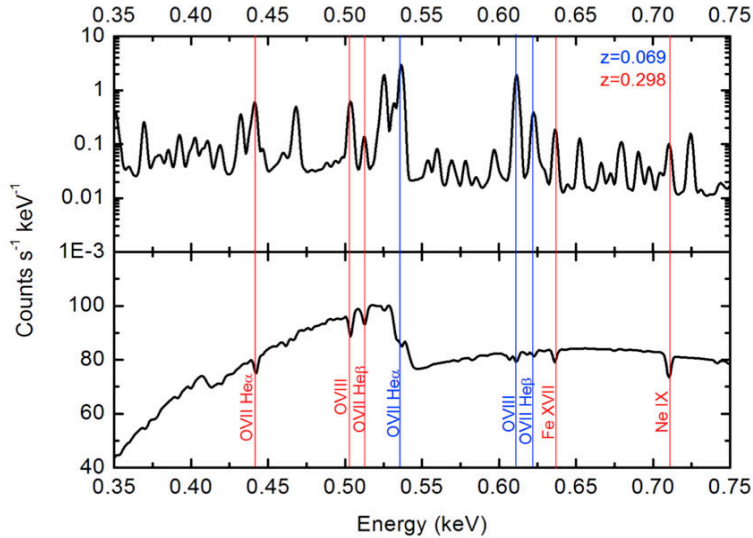

Figure 5: **Simulated emission and absorption line spectra captured in a single *Athena+* observation for two filaments at different redshifts.** Lower panel: absorption spectrum from a sight line where two different filamentary systems are illuminated by a bright background source. Upper Panel: corresponding emission from a 2'x2' region from the same filaments for a 1 Ms exposure time. The high spectral resolution allows us to distinguish both components. *Athena+* will be able to study ~100 of these sight lines in detail.

X-rays. Present facilities can marginally detect a few filaments (Nicastro et al 2013), but not characterize their physical properties. *Athena+* will probe these baryons in three dimensions, through a combination of absorption and emission studies using the X-IFU. Deep observations of bright AGN combined with Gamma-Ray Burst (GRB) afterglows caught with a 2-4 hour reaction time will be used as backlights for absorption studies through the warm and hot gas. Lines from the high ionization states of O, Ne, Si and Fe, seen simultaneously, enable unique identification of the filamentary structures of the cosmic web (Figure 5), with the detection and characterization of about hundred filaments. At the same time the emission from these structures is mapped by X-ray lines. Combining the two measurements allows the projected size of the structures to be derived while the shapes of the lines and their position reveal the kinematics of the baryons, which, together with the clustering information from the emission lines, pinpoints their origin for the first time.

In Table 1, we summarize the key issues addressed in this section.

**Table 1: The Hot Universe: Key issues and key observations.**

| How does ordinary matter assemble into the large scale structures that we see today? | |
|---|---|
| Key issue | *Athena+* key observation |
| The formation and evolution of groups and clusters of galaxies | |
| Understand how baryons accrete and evolve in the largest dark matter potential wells of groups and clusters. Determine how and when the energy contained in the hot intra-cluster medium was generated. | Map the structure of the hot gas trapped in galaxy clusters at various redshifts out to the virial radius, resolving gas density and temperature with the WFI. Measure the gas motions and turbulence through X-IFU spatially resolved spectroscopy. |
| The chemical history of the hot baryons | |
| Determine when the largest baryon reservoirs in galaxy clusters were chemically enriched. Infer the relative contributions of supernova types, and the initial stellar mass function in protoclusters. Identify the locations in clusters where most of the metals are generated, and determine how they are dispersed. | Measure elemental abundances of heavy elements like O, Ne, Mg, Si, S and Fe, through X-IFU X-ray spectroscopy of groups and clusters at different redshifts. Synthesize the abundances using yields of various SN types and AGB stars. Determine where metals are produced in clusters via spatially resolved spectroscopy of nearby objects. |
| Cluster feedback | |
| Understand how jets from AGN dissipate their mechanical energy in the intracluster medium, and how this affects the hot gas distribution. | Measure hot gas bulk motions and energy stored in turbulence directly associated with the expanding radio lobes in the innermost parts of nearby clusters with X-IFU. Use sensitive WFI imaging to detect and characterize large scale ripples and weak shocks in nearby groups and clusters. |
| Determine whether jets from powerful radio-loud AGN are the dominant non-gravitational process affecting the evolution of hot gas in galaxy groups and clusters. | Use WFI to obtain temperature maps of clusters around radio-loud AGN out to intermediate redshifts and map shock structures. Test jet evolution models and infer their impact at the epoch of group and cluster formation. |
| Establish how AGN feedback regulates gas cooling in groups and clusters and AGN fuelling | Compare jet power estimates by determining total energy budget and dynamical timescales from X-IFU velocity measurements, with accretion rates for competing fuelling models tuned to precisely measured thermodynamical conditions. Determine importance of AGN-induced turbulence in driving thermal instabilities, via mapping of turbulent velocities in a range of systems. |
| The Warm-Hot Intergalactic Medium | |
| Find the missing 50% of baryons at $z<2$ and reveal the underlying mechanisms driving the distribution of this gas on various scales, from galaxies to galaxy clusters, as well as metal circulation and feedback processes. | Determine the distribution of filaments via X-ray absorption spectroscopy against bright distant objects with X-IFU. For the fraction that can be seen in emission, measure their chemical composition, density, size, temperature, ionization and turbulence. |

## 2.2. The Energetic Universe: how do black holes grow and influence the Universe?

All massive galaxies, not just those in clusters and groups, host a supermassive black hole (SMBH) at their centre, the mass of which is tightly correlated with the galaxy bulge properties (e.g. via the $M_{BH}$-$\sigma$ relation). This observation has revolutionized our view of the formation and evolution of galaxies, implying a profound influence of black hole accretion throughout the Universe (Kormendy & Ho, 2013). The energy released during the build up of the SMBH exceeds the binding energy of the entire galaxy by a factor of 10-100, but the relationship requires a self-regulating mechanism connecting the accretion-powered growth of the SMBH at the event horizon level to the star-formation powered growth of the galaxy at much larger scales. Determining the nature and prevalence of this feedback is key to understand the growth and co-evolution of black holes and their host galaxies. This **Energetic Universe** is revealed in a unique manner by observations in the X-ray band, as this is where contamination from the host galaxy is smaller. X-rays provide the clearest and most robust way of performing a census of black hole growth in the Universe, accounting for obscured objects. Despite the progress made currently by *Chandra* and *XMM-Newton*, order of magnitude increases in both survey power and photon collecting area are required to address the most pressing issues in the global black hole growth history, namely its evolution at high redshift ($z$=4-10) and the importance of ultra-obscured, Compton-thick objects. On smaller scales, X-rays produced by gravitational release near the event horizon of black holes are able to diagnose the accretion flow in the strong gravity regime. These observations are needed to reveal not only the secrets of how matter flows on to the black hole, but how and why the various types of outflow (jets and winds) are launched. The capabilities to probe both the event-horizon scale and the most distant AGN are provided by *Athena+*, offering a unique and comprehensive view of **The Energetic Universe**. With these capabilities, we will address the following questions:

- How do early supermassive black holes form, evolve and affect the distant Universe?
- What is the role of (obscured) black hole growth in the evolution of galaxies?
- How do accretion-powered outflows affect larger scales via feedback?
- How do accretion and ejection processes operate in the near environment of black holes?

*Athena+* will explore many aspects of black hole accretion and provide definitive evidence of how black holes grow, where in the Universe that growth occurs, and how it affects the wider cosmos.

### 2.2.1. Formation and Early Growth of Supermassive Black Holes

The processes responsible for the early growth of SMBH are currently unknown. The remnants of the first generation of stars (PopIII stars) may be the seeds of SMBHs and must grow rapidly through frequent periods of intense accretion in early galaxies (Li et al. 2007). Alternatively, massive seeds may form from the monolithic collapse of primordial gas clouds (Begelman et al. 2006), and grow through extended periods of more moderate accretion (Figure 6). How these early AGN shape the evolution of their host galaxy via feedback and the role they may play in the reionisation of the Universe is also unknown. Rapid growth requires considerable fuel for accretion so is likely to imply obscuration. Current and future facilities at (e.g. ALMA, JWST, E-ELT) have a strong focus on finding and studying galaxies, the high redshift Universe, and may sometimes reveal spectroscopically whether these galaxies have a growing black hole. To perform a census, however, requires a selection technique which traces the bolometric luminosity of the system, while minimizing the effects of obscuration and contamination from star formation processes. X-rays provide this, and hence the most powerful AGN selection method at high redshift. We must also cover very large sky areas at very high sensitivity, something that current X-ray instruments (and the above-mentioned longer wavelength facilities), cannot do.

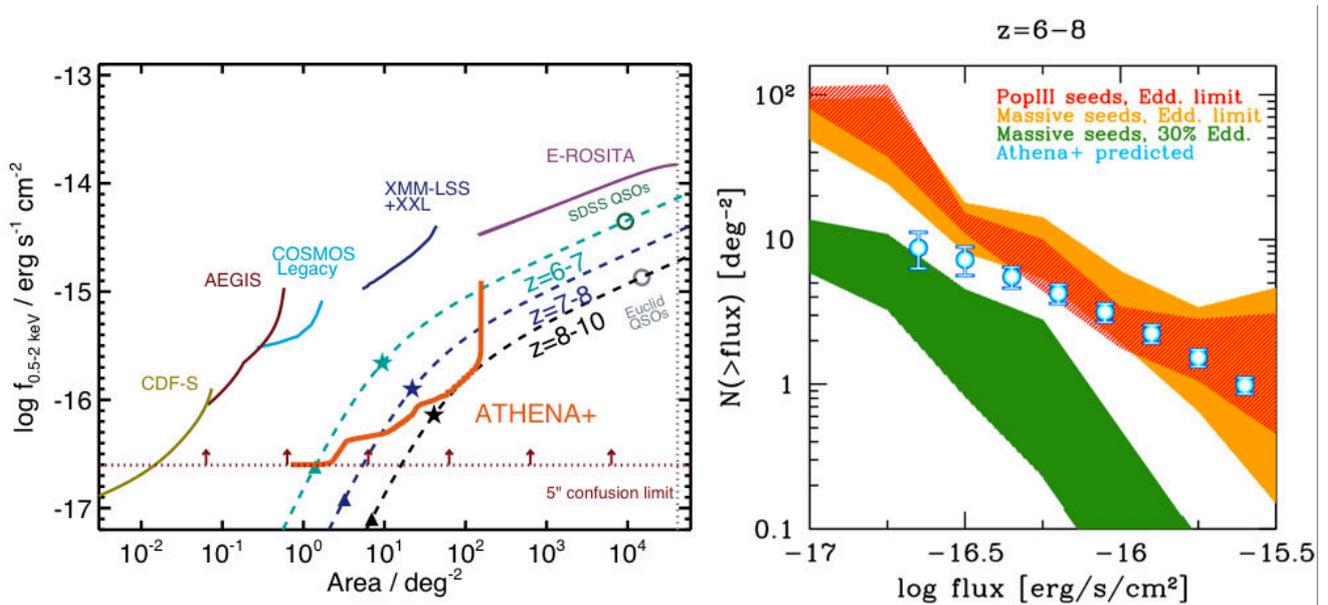

Figure 6: **The new discovery space that will be accessed by *Athena+* surveys, which will place essential constraints on physical models for early SMBH growth and formation.** *Left*: Area-flux coverage for a multi-tiered survey with the *Athena+* WFI (red line; 4x1Ms, 20x300ks, 75x100ks, 250x30ks), compared with existing and planned X-ray surveys. *Athena+* will break through to the high redshift Universe for the first time, with a survey power a factor ~100 better then existing surveys (other solid lines). The dashed lines show the coverage required to detect at least ten sources in the given redshift ranges. *Athena+* will probe the population ~2-3 orders of magnitude fainter than the SDSS or Euclid QSOs (open circles), reaching the population of objects which dominate the accretion power. *Right*: Expected number counts of $z$=6-8 AGN from the survey (circles). Note that at present no purely X-ray selected objects have yet been found in this redshift range. The shaded regions show predictions based on theoretical models that differ by black hole formation mechanism and growth rate (Volonteri & Begelman, 2010).

Despite large time investments, current X-ray facilities have been able to determine AGN demographics only to around $z\sim4$ (Aird et al. 2010). With its combination of collecting area at ~1 keV, field-of-view and angular resolution, *Athena+* will perform X-ray surveys more than two orders of magnitude faster than *Chandra* or *XMM-Newton*. *Athena+* will access new discovery space, so that models of seed formation and growth mechanisms for early SMBHs in the $z$=6-10 redshift range can be discriminated. Currently there is no X-ray discovered object confirmed above $z$=6, i.e. into the epoch of reionization where the first galaxies and black holes formed. A suitably designed multi-tiered survey with the *Athena+* WFI will yield over 400 $z = 6-8$ X-ray selected AGN and around 30 at $z$>8 (Figure 6). Crucially, these will include obscured objects, and those whose host galaxy light prevents identification of the AGN in other wavebands. Note that performing an equivalent survey with *Chandra*, the only current X-ray facility sufficiently sensitive to access the necessary flux range, would take approximately 100 years of dedicated survey observations.

*Athena+* surveys will also pinpoint active SMBHs (to ~1" positional accuracy) within samples of galaxies from the large, state-of-the-art optical and near-infrared imaging surveys (e.g. LSST, Hyper Suprime Cam, Euclid). Further follow-up of these X-ray AGN with ALMA, E-ELT and JWST will yield redshifts, stellar masses, star formation rates, cold gas properties, dust masses, and other important properties of the host galaxies. *Athena+* X-IFU follow-up of samples of $z\sim6$-9 galaxies may even reveal the intense iron line emission using the superb spectral resolution to improve the sensitivity and push below the confusion limit. This emission characterizes the most obscured AGN, which may signpost a critical phase in the formation of the earliest galaxies.

An entirely complementary way of exploring the seed population of SMBH at the highest redshifts can be achieved via fast follow-up observations of X-ray afterglows of Gamma-Ray Bursts. The chemical fingerprint of Pop III star explosions is distinct from that of later generations (Heger & Woosley 2010). Tracing the first generation of stars is also crucial for understanding cosmic re-ionization and the dissemination of the first metals in the Universe. High resolution absorption line X-ray spectroscopy with the *Athena+* X-IFU will enable us to determine the redshift and the typical masses of early stars, with fundamental impact on the models of early star populations and the onset of the accretion power from SMBH in the Universe.

### 2.2.2.  Obscured Accretion and Galaxy Evolution

Following their formation at high redshifts, galaxies and black holes continue to grow, with the peak of both star formation and accretion activity occurring at $z\sim1$-4. The general picture proposed involves an early phase of intense black hole growth that takes place behind large obscuring columns of inflowing dust and gas clouds. This is then followed by a stage in which some form of AGN feedback controls the fate of the interstellar medium and hence, the evolution of the galaxy. Open questions that relate to our current understanding of black hole growth and its relation to the build-up of galaxies include: what are the physical conditions (e.g. fuelling mode, triggering mechanism) that initiate major black hole accretion events and how they are connected to star-formation on larger scales; what is the nature of AGN feedback and does it play a significant role in the evolution of galaxies in the range $z\sim1$-4, where most of the stellar and black hole mass was put in place.

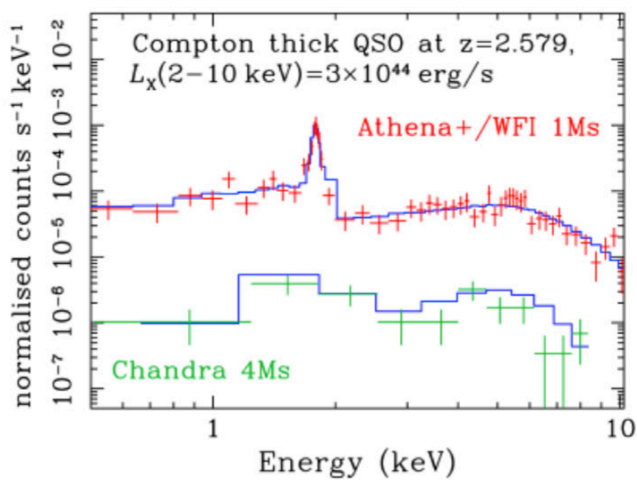

Figure 7:  ***Athena+* WFI spectrum of a Compton thick AGN at $z$=2.6.** The spectrum accumulated by *Chandra* in 4 Ms is shown for comparison.

Observations at high energies provide unique information about both the early heavily obscured stage and the later blow-out phase. Detection of the intense iron K$\alpha$ emission (Figures 1 and 7), characteristic of the most obscured "Compton thick" phase, will yield clean samples of deeply shrouded AGN with well-defined selection functions and determinations of their accretion luminosity and obscuring column. Existing X-ray surveys have yielded only a few tens of the most obscured sources. *Athena+* WFI surveys will yield around 15,000 of the most heavily obscured AGN to $z\sim4$, and perhaps beyond (Figure 1), determining their demographics for the first time.

X-rays also probe the most highly ionized component of feedback flows from AGN, which can dominate in terms of energy and mass flux, by detecting highly ionised material via absorption line spectroscopy (Figure 1). This material is invisible at other wavelengths. Current studies of these phenomena are largely limited to the local Universe. The high throughput of *Athena+* will enable detailed X-ray spectral studies of AGN up to $z\sim4$, including the prevalence and energetics of these feedback flows, for the first time.

Having identified the key evolutionary stage via X-ray spectroscopy, future observatories that focus primarily on galaxy properties will provide the necessary complementary information (e.g. morphology, star formation) to understand the link between black hole growth and galaxy formation when the Universe was experiencing its most active phase.

The same surveys, designed to yield significant samples of relatively rare populations (e.g. high $z$ or Compton thick AGN), will also yield of order 600,000 mildly and unobscured AGN at all redshifts. This will enable detailed statistical investigations, e.g. of AGN host galaxies, clustering, and the link between black hole accretion and large scale structures as a function of redshift, luminosity or obscuration. The huge improvement in sample sizes will open up a new discovery space in supermassive black hole studies, similar to the progress achieved in galaxy evolution work when huge datasets (e.g. SDSS) became available (e.g. the discovery of star-formation main sequence, galaxy colour bimodality). As an example, we note that the baseline *Athena+* WFI survey strategy will yield approximately 10,000 X-ray AGN at $z$=4-6, compared to about 100 current examples from combined *Chandra* and *XMM-Newton* surveys.

### 2.2.3.   Galaxy-scale Feedback

While AGN feedback is invoked in almost all models of galaxy evolution, the physical mechanisms by which the energy output from the AGN emerges and couples with the surrounding medium at larger scales is not yet established. Jet feedback is well established in clusters of galaxies, where it heats gas in the halo preventing gas cooling onto the central galaxy (Croton et al. 2006, see Fabian 2012 for a recent review). The AGN feedback mechanism that is thought to quench star formation in more typical massive galaxies is likely to be different. In this context, the high velocity winds recently discovered in the X-ray spectra of AGN are potentially a very effective way of transporting energy from the nuclear scale to galaxy (King & Pounds, 2003). These are important, because they probe the phase of the wind/outflow which carries most of the kinetic energy, and is otherwise too highly ionized to be seen at longer wavelengths. According to existing models, the energy of such powerful AGN-driven winds is deposited into the host galaxy ISM, and may contribute to the powering of the recently discovered galactic-scale molecular outflows, which are able to sweep away the galaxy's reservoir of gas and quench the star formation activity (Wagner et al. 2013).

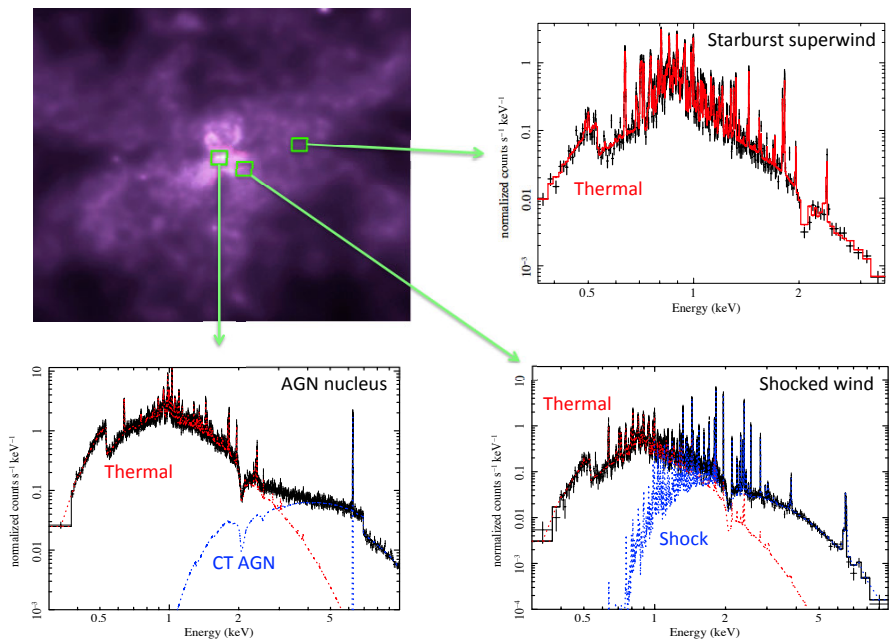

Figure 8: **Simulated *Athena+* X-IFU spectra of different regions in the ULIRG NGC 6240, illustrating the ability to disentangle the complex mix of excitation mechanisms**. The observations can clearly distinguish starburst-driven excitation and AGN-heated shocks in the large scale galactic outflows of molecular and cold gas, as well as the buried component originating from the Compton thick AGN double nuclei. The top left panel shows the *Chandra* X-ray image of NGC6240 (Credit: NASA/CXC/SAO/E. Nardini et al.), with the green boxes indicating the representative 5"x5" regions for which the spectra are simulated.

*Athena+* is sufficiently sensitive to detect these winds at cosmological distances. In the local Universe, exquisite detail of the feedback process will be revealed. The absorption lines from AGN winds and outflows and, crucially, their variability will determine the ionization state, density, temperature, abundances, velocities and geometry of the winds down to the inner regions where they are launched (Proga 2000). We expect different correlations between the physical parameters (e.g. density, ionization parameter) as a function of distance, maximum outflow velocity, and X-ray or UV luminosity. These can then be compared to theoretical models of accretion disk winds.

With the spatially-resolved spectroscopic power of the X-IFU, the interactions between hot and ionized feedback winds – whether originating from AGN or star formation – can be resolved and distinguished (Figure 8). The ability to separate spatially and spectrally the two types of activity will distinguish the dominant feedback mode and reveal if and how these phenomena are linked. The *Athena+* X-IFU will clearly distinguish spectrally between starburst superwinds from AGN-driven shocks on scales down to few kpc in 40-50 nearby AGN, ULIRG and starburst galaxies. At the same time it will map the velocity field of the outflowing gas to 20-30 km s$^{-1}$. These results will provide a template for AGN at higher redshift where wind shocks can be spectrally, but not spatially, resolved.

Supernovae-driven hot winds are in themselves a crucial element in our understanding of galaxy evolution and baryonic structures. Hydrodynamical simulations show that they are able to expel 90% (or more) of the baryons in star-forming galaxies, transporting metals into the intragroup, intracluster or inter-galactic medium. *Athena+* will observe this process directly via spatially resolved spectroscopy of nearby galaxies, and in particular star-forming galaxies. Currently, we have crude constraints on the X-ray halo of either very nearby and bright examples (i.e. M82) or very massive galaxies through stacked analysis. *Athena+* will perform high-resolution spectroscopy on typical galaxies with high spatial resolution, revealing how much mass and metals are carried in the hot superwind component.

### 2.2.4. The Physics of Accretion

The black hole feedback processes that shape the evolution of galaxies ultimately originate on scales 10 orders of magnitude smaller, as a result of accretion processes close to the event horizon. To discover this mechanism we must access a very small region, within a few tens of gravitational radii of the compact object. It is here where matter in the accretion disk may release almost half of its rest-mass energy, where winds and jets are launched, and where General and Special Relativity leave their mark on the emitted radiation. To understand accretion we must therefore understand the close environment of the black hole, how strong gravity affects the behaviour of matter and radiation, whether or not the black hole is spinning, and the relationship between accretion and ejection. Important questions still remain, such as whether the disk always extends down to the innermost stable orbit and, if not, whether any disk truncation is related to the launching of winds and jets. Similarly, how the jets themselves are accelerated remains largely a mystery, with wound-up magnetic fields usually invoked. It is also possible that black hole spin is tapped by a variant of the Penrose process (Blandford & Znajek, 1977).

X-ray emission is produced copiously in the black hole environment, via Comptonisation of thermal disk photons by electrons in a hot corona. The resulting continuum illuminates the disk, where it is reprocessed and scattered, producing signatures in the X-ray spectrum such as the iron Kα line which encode the signatures of the strong gravity environment (Nandra et al. 2007). Spectroscopy and timing of these reflection features is the key to further understanding of the accretion process. The most promising technique involves mapping of the inner accretion flow via the reverberation expected when intrinsic changes in the luminosity of the corona are seen in the reflection spectrum (Stella 1990). Differing path lengths mean that the reflection is seen to lag behind the coronal continuum. Since different parts of the reflection spectrum come from different radii of the disc, the lag energy spectrum depends on both time and energy and can be used to determine both the mass and spin of the black hole, and to map the central regions. The predicted lags have now been measured by *XMM-Newton* (Fabian et al. 2009) and confirmed by microlensing observations, placing limits on the size of the corona to a few tens of gravitational radii.

*Athena+* will take the next step, improving the data quality sufficiently to determine the so-called transfer function of the process (Blandford & McKee 1982), which encodes the underlying geometry of the corona-disk system. For example, it will be possible to distinguish between a compact corona on the black hole spin axis (suggestive of e.g. an aborted jet) or one extended over the disk. *Athena+* will reverberation-map hundreds of AGN and many black hole binary systems (and also neutron star binaries), exploiting the large effective area, angular resolution, spectral resolution, timing capabilities, soft energy response and uninterrupted long exposures provided by the mission and its orbit. This will enable detailed exploration and calibration of a wide range of effects in the brighter AGN, in particular in objects with jets or winds where it may reveal the important acceleration zone. The behaviour of the inner accretion flow where the energy release occurs will be revealed and, from a large sample of both massive and stellar mass black holes, show how and why the gravitational energy is split between the disk emission, the hot corona, the fast outflow and a jet, the last three presumed to be magnetically powered.

Reverberation signals also allow measurement of the black hole spin. In fact, *Athena+* will offer multiple probes of spin, such as disk continuum fitting and aperiodic variability in black hole binaries. These measurements in bright sources will allow calibration of the most common method, which exploits the shape of the broad iron Kα line (Fabian et al. 2000). This will then open up these measurements at cosmologically significant distances ($z$=1-2), with the iron line region redshifted to the 1-3 keV range where *Athena+* has enormous effective area. This could enable a unique probe of the black hole evolution, via the determination of the spin distribution in AGN, which encodes information about whether the black holes grew mainly from accretion or mergers (Berti & Volonteri 2008). In Table 2, we summarize the key issues for the hot Universe.

**Table 2: The Energetic Universe: key issues and key observations.**

| The Energetic Universe: how do black holes grow and influence the Universe? | |
|---|---|
| **Key issue** | *Athena+* **key observation** |
| **Formation and early growth of supermassive black holes** | |
| Determine the nature of the seeds of high redshift ($z$>6) SMBH, which processes dominated their early growth, and the influence of accreting SMBH on the formation of galaxies in the early Universe. | Accreting SMBH, even in obscured environments, will be detected out to the highest redshifts through their X-ray emission in multi-tiered WFI X-ray surveys. The most obscured objects will be unveiled by targeted X-IFU spectroscopy revealing strong reflected iron lines. |
| Trace the first generation of stars to understand cosmic re-ionization, the formation of the first seed black holes, and the dissemination of the first metals. | X-IFU measurements of metal abundance patterns for a variety of ions (e.g., S, Si, Fe) for at least 10 medium-bright gamma-ray burst X-ray afterglows per year with H equivalent column densities as small as $10^{21}$ cm$^{-2}$ and gas metallicities as low as 1% of solar. |
| **Obscured accretion and galaxy formation** | |
| Find the physical conditions under which SMBH grew at the epoch when most of the accretion and star formation in the Universe occurred ($z$~1-4). | Perform a complete census of AGN out to $z$~3, including those that reside inside a Compton-thick environment. This will be achieved via WFI surveys, where strong iron lines will be the signposts of heavily obscured AGN. |
| **Galaxy-scale feedback** | |
| Understand how accretion disks around black holes launch winds and outflows and determine how much energy these carry. | Use X-IFU to fully characterize ejecta, by measuring ionization state, density, temperature, abundances, velocities and geometry of absorption and emission features produced by the winds and outflows in tens of nearby AGN. |
| Understand the significance of AGN outflows in determining the build-up of galaxies at the epoch when most stars in present day galaxies formed. | X-IFU observations of nearby AGN/ULIRGs/starbursts will probe the interactions of AGN- and starburst-driven outflows with the ISM, and will provide a local template for understanding AGN feedback at higher redshift. |
| Understand how the energy and metals are accelerated in galactic winds and outflows and are deposited in the circum-galactic medium. Determine whether the baryons and metals missing in galaxies since $z$~3 reside in such extended hot envelopes. | Use X-IFU to directly map galactic haloes in nearby galaxies to characterize warm and hot gas outflows around starburst, ULIRG and AGN galaxies. Measure gas mass deposited, mechanical energy and chemical abundances to model baryon and metal loss in galaxies across cosmic time. |
| **The physics of accretion** | |
| Determine the relationship between the accretion disk around black holes and its hot electron plasma. Understand the interplay of the disk/corona system with matter ejected in the form of winds and outflows. | Perform time-resolved X-ray spectroscopy of X-ray binaries and AGN out to significant redshifts ($z$~1), so that time lags between different spectral components can be found and the transfer function measured. These measurements will then determine the geometry of the disk/corona system, key to understanding how jets and winds are launched. |
| Infer whether accretion or mergers drive the growth of SMBH across cosmic time. | Measure black hole spins through reverberation, timing, time-resolved spectroscopy and average spectral methods. Use spectroscopy to perform a survey of SMBH spins out to $z$~1-2 and compare with predictions from merger and accretion models. |

## 2.3. Science enabled by the *Athena+* observatory capabilities

The **Hot and Energetic Universe** includes almost all known astrophysical objects, from the closest planets to the most distant quasars and gamma-ray bursts. The instrument suite required to achieve the science goals described above provides *Athena+* with unprecedented observatory capabilities, enabling new science to be performed for a wide range of objects, of great interest to the whole astronomical community. Here, we provide a non-exhaustive list of scientific issues which can be addressed by *Athena+,* together with a short description of the *Athena+* breakthrough observations to be performed (Table 3).

**Table 3: Science enabled by the observatory capabilities of *Athena+***

| Key issue | *Athena+* key observation |
|---|---|
| **Planets** | |
| Establish how planetary magnetospheres and exospheres, and comets, respond to the interaction with the solar wind, in a global way that in situ observations cannot offer. | First detailed spectral mapping of Jupiter's X-ray emission, of the Io Plasma Torus, of Mars' exosphere and of X-rays from comets. Fluorescence spectra of Galilean Satellites for surface composition analysis. |
| **Exoplanets** | |
| Extend exoplanet research to incorporate X-ray studies to explore the magnetic interplay between stars and planets. | Measurements of X-ray spectral variability over the activity cycle of the host star and over the planet's orbital period. |
| **Stellar physics** | |
| Assess the mass loss rates of high velocity chemically-enriched material from massive stars to understand the role they play in the feedback processes on Galactic scales. | Time-resolved X-IFU spectroscopy of single and binary massive stars to characterize the large scale structures in their winds and assess their mass-loss rates. |
| Understand how high-energy irradiation of disks during the formation and early evolution of low and intermediate-mass stars affects disk evolution and eventually planetary system formation | Time resolved X-IFU spectroscopy of the brightest objects to explore the accretion process variability and the modulation due to accretion stream shadowing, and constrain the bulk velocity of accreting material with Doppler line shifts. |
| **Supernovae** | |
| Understanding the physics of core collapse and type Ia supernova remnants, quantifying the level of asymmetry in the explosion mechanism, the production of heavy elements, and their impact on the galactic environment. | First detailed 3D mapping of the hot ejected material in the line of sight (velocity, temperature, ionization state and composition) to determine to the full geometry and properties of the different layers of shocked ejecta. |
| **Stellar endpoints** | |
| Discover how mass loss from disk winds influences the binary evolution and impact the interstellar medium? | Perform multiple X-IFU observations on time-scales shorter than the wind time variability, measuring velocities and ionisation states in the outflow. |
| Extending the measurements of mass and radius of neutron stars to isolated millisecond pulsars and faint quiescent neutron star binaries | Waveform fitting of X-ray pulses from isolated millisecond pulsars and modeling of atmospheric emission from globular cluster sources. |
| **Sgr A\*** | |
| Understand flare production in Sgr A\*, the origin of the quiescent emission, and set constraints on the past AGN activity of Sgr A\*. | X-IFU observations along with multi-wavelength coverage to measure the ionization process and physical properties of the plasma during the flaring and quiescent states. |
| **Interstellar dust** | |
| Understand the chemical composition of interstellar dust. | X-IFU observation of extended X-ray absorption features. |
| **Interstellar medium** | |
| Determine the chemical composition of the hot gas of the interstellar medium, as a tracer of stellar activity in our and other galaxies. | X-IFU spectrum of the hottest emission and absorption components of the ionized gas characterized by e.g. OVII, OVIII, NeIX |

## 3.  SYNERGIES AND DISCOVERY SPACE

In the late 2020s, the pre-eminent facilities operating at other wavelengths will include LOFAR, SKA, ALMA, GAIA, JWST, E-ELT, LSST and CTA. *Athena+* has capabilities unrivalled by any other planned high-energy mission, complementing those facilities by providing the essential panchromatic view needed to understand a wide range of astrophysical phenomena. Figure 9 provides one illustration of this complementarity, focusing on the study of the first galaxies in cosmic history, emphasizing the excellent match between the sensitivity offered by *Athena+* and the anticipated capabilities at other wavelengths. This is necessary to understand the complex nature and phenomenology of the vast majority of astrophysical sources, where high-energy phenomena are present and X-ray observations therefore provide unique information.

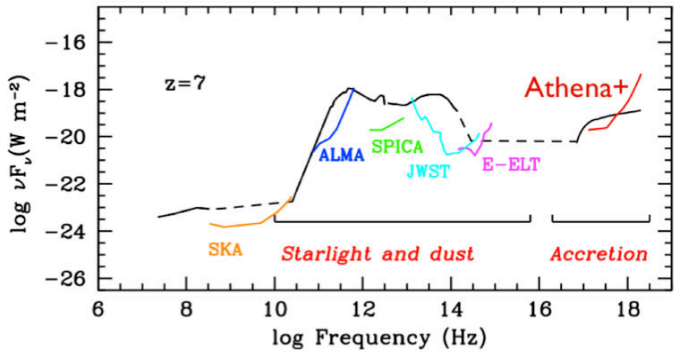

Figure 9: **Spectral Energy Distribution (SED) of a high redshift, obscured AGN at z=7.** The average SED, adapted from Lusso et al. 2011, is shown as the solid line. In terms of bolometric luminosity, this object is representative of what will be detected in Athena+ medium-deep surveys (e.g. 300 ks exposures, red line). The 3σ sensitivities (for a ~40ks exposure) of SKA, ALMA, SPICA, JWST and E-ELT are also shown, as labeled.

*Athena+* will deliver a rich return of serendipitous discoveries by opening up new discovery space, enabled by a major step forward in observational capabilities. Historically, high-energy astrophysics missions have been very effective in this respect. Some examples include the discovery of vast amounts of hot baryonic gas trapped in the potential wells of galaxy clusters and the clear demonstration of the existence of growing super-massive black holes in many galaxy cores previously believed to be normal systems. How *Athena+* will explore its discovery space cannot be stated in advance, but the observatory nature of the mission ensures that it will be driven by cutting-edge science questions. As an example, its unprecedented sensitivity will reveal populations of X-ray sources never seen before, in particular some of the first galaxy groups at $z\sim2.5$-3, predecessors of today's galaxy clusters.

## 4.  THE ATHENA+ MISSION CONCEPT

Achieving the ambitious goals set out in this White Paper requires an X-ray observatory-class mission delivering a major leap forward in high-energy observational capabilities. Thanks to its revolutionary optics technology and the most advanced X-ray instrumentation, the *Athena+* mission, outlined below, will deliver superior wide field X-ray imaging, timing and imaging spectroscopy capabilities, far beyond those of any existing or approved future facilities. Like *XMM-Newton* today, *Athena+* will play a central role in astrophysical investigations in the next decade. No other observatory-class X-ray facility is programmed for that timeframe, and therefore *Athena+* will provide our only view of **the Hot and Energetic Universe**, leaving a major legacy for the future. The *Athena+* mission has an exceptionally mature heritage based on extensive studies and developments by ESA and the member states for *Athena* (and IXO). Compared with *Athena,* the *Athena+* concept incorporates important enhancements, including a doubling of the effective area (to 2m²); an improvement in the angular resolution by a factor ~2  (to 5") and quadrupling of the fields of view of both the WFI and X-IFU, yet representing a realistic evolution in performance for a mission to fly in 2028.

Mapping the dynamics and chemical composition of hot gas in diffuse sources requires high spectral resolution imaging (2.5 eV resolution) with low background; this also optimizes the sensitivity for weak absorption and emission features needed for WHIM studies or for faint point source characterisation. An angular resolution of 5 arcsec is required to disentangle small structures of clusters and groups and, in combination with a large area, provide high resolution spectra, even for faint sources. This angular resolution, combined with the mirror effective area and large field of view (40 arcmin) of the WFI provides the detection sensitivity (limiting flux of $10^{-17}$ erg/cm²/s 0.5-2 keV band) required to detect AGNs at $z>6$ within a reasonable survey time. The science requirements and enabling technologies are summarized in Table 4.

**Table 4: Key parameters and requirements of the *Athena+* mission. The enabling technology is indicated.**

| Parameter | Requirements | Enabling technology/comments |
|---|---|---|
| **Effective Area** | 2 m² @ 1 keV (goal 2.5 m²)<br>0.25 m² @ 6 keV (goal 0.3 m²) | Silicon Pore Optics developed by ESA. Single telescope: 3 m outer diameter, 12 m fixed focal length. |
| **Angular Resolution** | 5" (goal 3") on-axis<br>10" at 25' radius | *Detailed analysis of error budget confirms that a performance of 5" HEW is feasible.* |
| **Energy Range** | 0.3-12 keV | Grazing incidence optics & detectors. |
| **Instrument Field of View** | *Wide-Field Imager:* (**WFI**): 40' (goal 50') | Large area DEPFET Active Pixel Sensors. |
| | *X-ray Integral Field Unit:* (**X-IFU**): 5' (goal 7') | Large array of multiplexed Transition Edge Sensors (TES) with 250 micron pixels. |
| **Spectral Resolution** | **WFI**: <150 eV @ 6 keV | Large area DEPFET Active Pixel Sensors. |
| | **X-IFU**: 2.5 eV @ 6 keV (goal 1.5 eV @ 1 keV) | *Inner array (10"x10") optimized for goal resolution at low energy (50 micron pixels).* |
| **Count Rate Capability** | > 1 Crab[3] (**WFI**) | *Central chip for high count rates without pile-up and with micro-second time resolution.* |
| | 10 mCrab, point source (**X-IFU**)<br>1 Crab (30% throughput) | *Filters and beam diffuser enable higher count rate capability with reduced spectral resolution.* |
| **TOO Response** | 4 hours (goal 2 hours) for 50% of time | *Slew times <2 hours feasible; total response time dependent on ground system issues.* |

## 4.1. Science payload

The strawman *Athena+* payload comprises three key elements:

- A single X-ray telescope with a focal length of 12m and an unprecedented effective area (2 m² at 1 keV). The X-ray telescope employs Silicon Pore Optics (SPO), an innovative technology that has been pioneered in Europe over the last decade mostly with ESA support. SPO is a highly modular concept, based on a set of compact individual mirror modules, which has an excellent effective area-to-mass ratio and can achieve high angular resolution (<5").
- The X-ray Integral Field Unit (X-IFU), an advanced actively shielded X-ray microcalorimeter spectrometer for high-resolution imaging, utilizing Transition Edge Sensors cooled to 50 mK.
- The Wide Field Imager (WFI), a Silicon Active Pixel Sensor camera with a large field of view, high count-rate capability and moderate resolution spectroscopic capability.

The two instruments (shown in Figure 10) can be moved in and out of the focal plane by an interchange mechanism, which is a simplified version of the IXO design. Key characteristics of the instruments are also listed in Table 4.

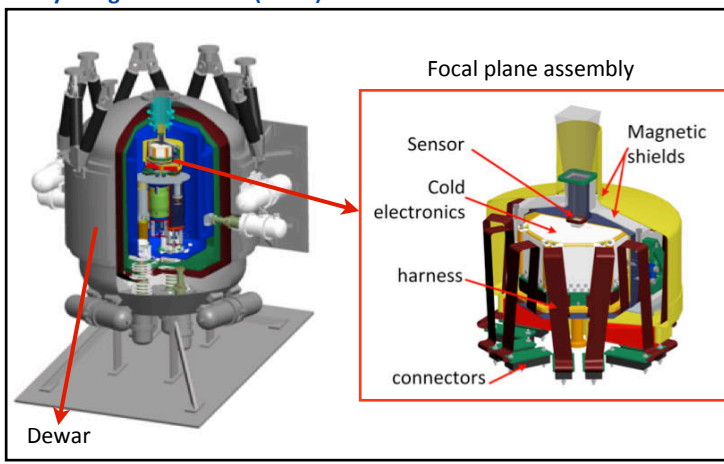
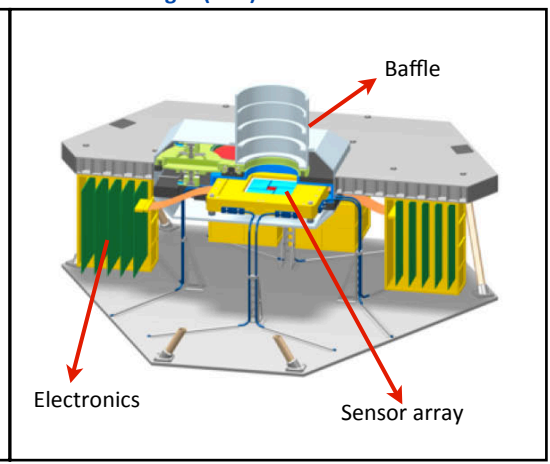

**X-ray Integral Field Unit (X-IFU)**　　　　　　　　　　　**Wide-Field Imager (WFI)**

Focal plane assembly — Sensor — Magnetic shields — Cold electronics — harness — connectors — Dewar — Electronics — Baffle — Sensor array

Figure 10: **The *Athena+* science instruments. *Left:*** Design drawing of the X-IFU showing the Dewar and a zoom on the focal plane assembly. ***Right:*** Design drawing of the WFI.

---

[3] 1 Crab corresponds to a flux of 2.4 10⁻⁹ ergs/s/cm² (2-10 keV).

## 4.2. Athena+ performance

The *Athena+* telescope delivers a throughput a factor ~10 larger than *XMM-Newton* and almost a factor 100 larger than *Astro-H* at low energies (and more than a factor 10 larger at high energies), coupled with major improvements in focal plane instrumentation, including the use of a large format microcalorimeter to provide high-resolution spectroscopic imaging and an advanced Si-sensor to provide wide-field imaging with spectroscopic capability and a combination of high time resolution and count-rate capability. We have selected a few key comparisons to illustrate *Athena+* performance in Figure 11.

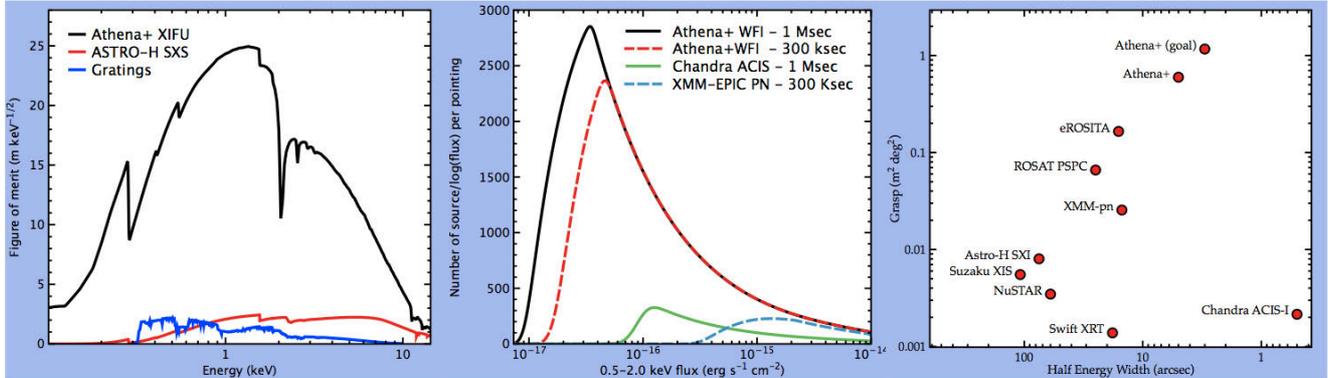

Figure 11: *Athena+ scientific performance. Left:* Figure of merit for weak spectral line detection of X-ray high-spectral resolution spectrometers, derived from the number of counts per independent spectral bin. The gratings line represents the best of the current *XMM-Newton* or *Chandra* grating at each energy. *Centre:* Number of sources per logarithmic flux interval expected in single *Athena+* WFI pointings at high Galactic latitudes compared to *Chandra* and *XMM-Newton*. *Right:* Grasp of previous, operational and planned missions as a function of angular resolution. Grasp is defined as the product of effective area at 1 keV (10 keV for NuSTAR) and the instrument field of view.

## 4.3. Mission profile

Preliminary industrial designs for the *Athena+* spacecraft are shown in Figure 12. Like *Athena*, it is a conventional design retaining much heritage from *XMM-Newton*. Considerations of observing efficiency and thermal stability favour an L2 orbit reached by Ariane V. The initial industrial assessments performed for this White Paper indicate very safe mass and power budget margins.

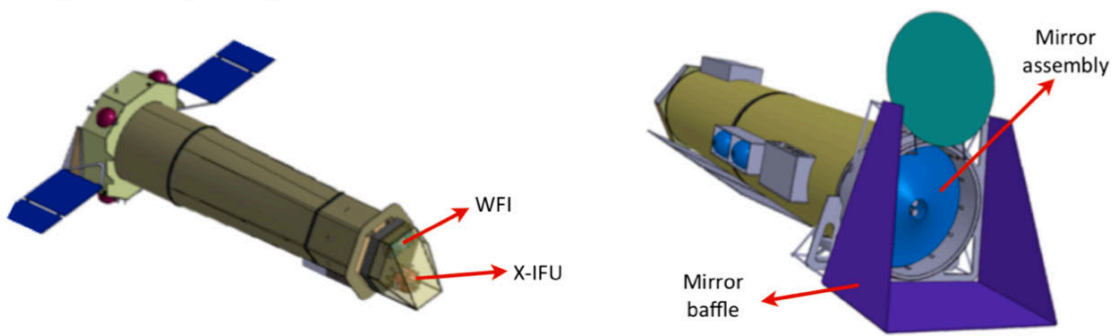

Figure 12: *Athena+ spacecraft designs. Left: Athena+* Astrium-UK satellite designs provided in the context of this White Paper. *Right:* Same for Thales-Alenia. For the Astrium design the interchangeable instruments are shown at the bottom, whilst the optics module is at the top and the solar panels will be unfolded. For the Thales-Alenia design the solar panels are body mounted and the optics module, with its unfolded cover and sunshield are visible.

Mission and science operations are conventional with community-based instrument and science data centre teams providing further support to ESA, as required. *Athena+* is an observatory whose program will be largely driven by calls for proposals from the scientific community, but may be complemented by key programs for science goals requiring large time investments. A nominal mission lifetime of 5 years would allow the core science goals set out in this White Paper to be achieved, while preserving a large fraction of the available time for broad based science programs.

Strong international interest in the mission has been expressed, based on the earlier collaborations with Japan (JAXA) and the US (NASA/GSFC, NIST) for *IXO* and *Athena*. These contributions could potentially reduce the costs to ESA and/or the ESA Member States, but *Athena+* can be implemented independently by Europe alone.

## 5. ACKNOWLEDGMENTS

We gratefully acknowledge the comments and inputs of the White Paper Review Team: M. Arnaud, J. Bregman, F. Combes, R. Kennicutt, R. Maiolino, R. Mushotzky, T. Ohashi, K. Pounds, C. Reynolds, H. Röttgering, M. Rowan-Robinson, C. Turon and G. Zamorani

## 6. REFERENCES

**Working Groups and Supporting Papers:** As part of the process of developing this White Paper we set up 13 Working Groups to look in detail at various science areas within the theme; much of the work of these groups is of course reflected in this document. The more detailed reports from the Working Groups provide more extensive discussion of the science topics and additional simulations and references. These reports, which we refer to as Supporting Papers, are available at http://www.the-athena-x-ray-observatory.eu (see Publications).